\documentclass[twocolumn,showpacs,preprintnumbers,amsmath,amssymb]{revtex4}
\usepackage{graphicx}
\usepackage{dcolumn}
\usepackage{bm}

\begin{document}

\title{Spin injection into a short DNA chain}
\author{X. F. Wang and Tapash Chakraborty}
\affiliation{Department of Physics and Astronomy, The University of Manitoba,
Winnipeg, Canada, R3T 2N2}

\begin{abstract}
Quantum spin transport through a short DNA chain connected to ferromagnetic
electrodes has been investigated by the transfer matrix method. We describe
the system by a tight-binding model where the parameters are extracted from
the experimental data and realistic metal energy bands. For ferromagnetic
iron electrodes, the magnetoresistance of a 30-basepair Poly(G)-Poly(C)
DNA is found to be lower than 10\% at a bias of $< 4$ V, but can reach
up to 20\% at a bias of 5 V. In the presence of the spin-flip mechanism,
the magnetoresistance is significantly enhanced when the spin-flip
coupling is weak but as the coupling becomes stronger the
decreasing magnetoresistance develops an oscillatory behavior.
\end{abstract}
\pacs{87.14.Gg,72.20.Ee,72.25.-b}
\maketitle

In recent years, a remarkable progress in direct measurements of electron
transport through DNA has generated intense interest in DNA electronics
\cite{review,pora1,yoo,carp,roch,pora2,endr,hjor,li,zwol,cuni,apal,bera}. DNA is
found to have diverse electronic properties depending on its structure and
the environment around it \cite{pora2}. The clear semiconductor behavior
observed in a short DNA chain of 30-basepair Poly(G)-Poly(C)
has been explained by a tight-binding model \cite{pora1,li,cuni}. On
the other hand, spin transport through nanostructures has, of late,
been receiving considerable attention because of the possiblity of
developing spin-based electronic devices \cite{spintro}. Inspired by
the broad interest in spin-injection into mesoscopic systems \cite{hu},
we have investigated the quantum spin transport \cite{zwol} through a
short DNA chain connected to ferromagnetic electrodes. We predict an
enhancement and oscillation of magnetoresistance in this system taking
into account the realistic band structure of ferromagnetic Fe electrode
and a spin-flip mechanism.

\begin{figure}
\begin{center}
\begin{picture}(330,130)
\put(0,0){\includegraphics{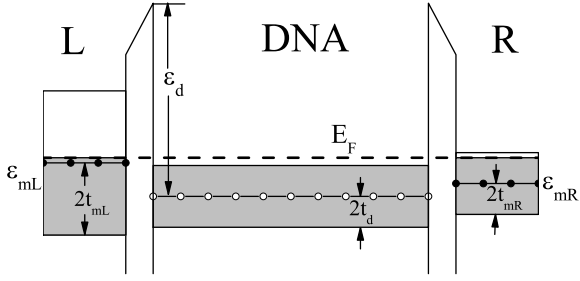}}
\end{picture}
\vspace{-1.0cm}
\protect\caption{Energy band of the system in equilibrium.}
\label{fig:fig1}
\end{center}
\end{figure}

We consider a $p$-type semiconductor DNA chain of $N$ basepairs connected to a
circuit via metal electrodes. The tight-binding Hamiltonian of the system is
\begin{equation}
H=H_{d}+H_{L}+H_{R}+H_{dm}+H_{dps}+H_{sp}
\end{equation}
where
\begin{widetext}
\begin{eqnarray*}
H_{d} &=&-\sum_{\sigma ,n=1}^{N}\varepsilon _{d}^{\sigma }C_{n,\sigma
}^\dag C_{n,\sigma }-\sum_{\sigma ,n=1}^{N-1}t_{d}^{\sigma }(C_{n,\sigma
}^\dag C_{n+1,\sigma }+C_{n+1,\sigma }^\dag C_{n,\sigma }), \\
H_{L} &=&-\sum_{\sigma ,n\leq 0}\varepsilon _{mL}^{\sigma }C_{n,\sigma
}^\dag C_{n,\sigma }-\sum_{\sigma ,n\leq 0}t_{mL}^{\sigma }(C_{n,\sigma
}^\dag C_{n+1,\sigma }+C_{n+1,\sigma }^\dag C_{n,\sigma }), \\
H_{R} &=&-\sum_{\sigma ,n\geq N+1}\varepsilon _{mR}^{\sigma }C_{n,\sigma
}^\dag C_{n,\sigma }-\sum_{\sigma ,n\geq N+1}t_{mR}^{\sigma }(C_{n,\sigma
}^\dag C_{n+1,\sigma }+C_{n+1,\sigma }^\dag C_{n,\sigma }), \\
H_{dm} &=&-\sum_{\sigma }t_{dm}(C_{0,\sigma }^\dag C_{1,\sigma }+C_{1,\sigma
}^\dag C_{0,\sigma })-\sum_{\sigma }t_{dm}(C_{N,\sigma }^\dag C_{N+1,\sigma
}+C_{N+1,\sigma }^\dag C_{N,\sigma }), \\
H_{dps} &=&-\sum_{\sigma ,n=1}^{N}\Sigma _{n}^{\sigma }(E)C_{n,\sigma
}^\dag C_{n,\sigma }, \\
H_{sp} &=&-\sum_{\sigma ,n=1}^{N}t_{d}^{so}C_{n,\sigma }^\dag C_{n,\bar{\sigma}%
}-\sum_{\sigma ,n=1}^{N-1}t_{d}^{s}(C_{n,\sigma }^\dag C_{n+1,\bar{\sigma}%
}+C_{n+1,\bar{\sigma}}^\dag C_{n,\sigma }).
\end{eqnarray*}
\end{widetext}
Here $C_{n,\sigma}^\dag$ is the creation operator of electron with spin
$\sigma$ on site $n$ ($=1,\cdots,N$) of the DNA chain, the left
electrode ($n\leq 0$), and the right electrode ($n\geq N+1$).
$H_d$ describes electrons (holes) of spin $\sigma$ in the DNA chain
with the on-site energies $-\varepsilon _{d}^{\sigma}$
($\varepsilon_d^{\sigma }$), which is equal to the highest occupied
molecular orbit (HOMO) energy of each base-pair, and the hopping
parameters $t_d^{\sigma}$ between neighboring sites. The HOMO energy
band is then determined by $\varepsilon_d^{\sigma}$ and $t_d^{\sigma}$.
The Fermi energy $E_F$ in the $p$-type DNA chain locates between the
HOMO and lowest unoccupied molecular orbit (LUMO) bands and is closer
to the HOMO band edge. Experimental results have indicated that the
Fermi energy may vary from sample to sample \cite{pora1}. $H_X$ with
$X=L$ or $R$ denotes the Hamiltonian of electrons in the left electrode
($L$) or the right electrode ($R$). In the tight-binding model,
$\varepsilon_{mX}^{\sigma}$ is the center of the energy band where the
electrons are in the metals and $4t_{mX}^{\sigma}$ its band width. When
the DNA chain contacts to the metal electrodes, exchange of electrons
(holes) between the DNA chain and the electrodes becomes possible. In
equilibrium, as illustrated in Fig.~\ref{fig:fig1}, the Fermi energies
of the electrodes and of the DNA match and a tunnelling barrier forms
between them. The contact property is described by the tunnelling
parameter $t_{dm}$. When a bias voltage drop is applied over the
electrodes, distribution of the voltage drop or the potential profile
along the non-equilibrium system depends on the DNA chain property and
its contact with the metal electrodes. Since the free electron density
in the metals is much higher than that in DNA, we assume that the band
structure of the metal electrode is not changed by the applied bias.
When the contact between DNA and the electrodes is poor, the voltage
drop concentrates on the contact. In case of a perfect contact however,
the whole voltage drop should be applied mainly along the DNA chain. In
this letter, we assume the voltage drop is on the contact since it is
supported by the fit to the experimental result (see below).

In real world, a DNA chain is composed of two strands of bases with one
phosphate-sugar backbone connected to each strand. The backbones can affect
the on-site energy of electrons in the basepairs. Further, the environment
around the DNA chain may also play a role in the property of the electrons.
Here we use a reservoir of semi-infinite chain \cite{amat,datt,li} with a
energy band of width $4\gamma $ and a coupling of strength $\eta$ to each DNA
basepair site to mimic the effect of the backbones and the environment. As a
result, the on-site energy of each site in the DNA basepair is modified by a \
self energy $\Sigma_n^{\sigma}(E)$ which is energy dependent and is expressed as,
$$\Sigma _{n}^{\sigma }(E)=\frac{\eta^2}{E-\varepsilon_r-\Sigma_r}$$
with $\varepsilon_r$ being the on-site energy of the semi-infinite reservoir
chain which we assume to be equal to the DNA on-site energy and
$\Sigma_r=(E-\varepsilon_r)/2-i[\gamma^2-(E-\varepsilon_r)^2/4]^{1/2}$
the self energy of any reservoir site which is obtained self-consistently.
In what follows, we have used the values $\eta=0.1$ eV and $\gamma=5$ eV \cite{li}.
To study the spin relaxation for a possible spin injection, we introduce
the term $H_{sp}$ in the Hamiltonian to take into account the spin flip on-site
and between neighboring sites described by the parameters $t_d^{so}$ and $t_d^s$
respectively. The spin-flip along the DNA can be a result of spin-orbit
interaction, magnetic impurity in the backbone, or magnetic environment.

In order to evaluate the transport properties of the system, we have
employed the transfer matrix method \cite{roch,carp}. For an open
system, the secular equation of the system is expressed as a group of
infinite number of equations of the form,
\begin{eqnarray*}
t_{n-1,n}\Phi_{n-1}^{\sigma} &+&t_{n-1,n}^s
\Phi_{n-1}^{\bar{\sigma}}+(\varepsilon_{n}^\sigma+\Sigma _n^\sigma-E)
\Phi_n^\sigma
+t_n^{so}\Phi_n^{\bar{\sigma}}\\
&+&t_{n,n+1}\Phi_{n+1}^\sigma+t_{n,n+1}^s\Phi_{n+1}^{\bar{\sigma}}=0.\\
\end{eqnarray*}
The wave functions of sites $n+1$ and $n$ are related to those of sites
$n$ and $n-1$ by a transfer matrix $\hat{M}$,
\begin{equation}
\left( \begin{array}{c}
\Phi_{n+1}^+ \\
\Phi_{n+1}^- \\
\Phi_n^+ \\
\Phi_n^-
\end{array}\right) =\hat{M}
\left(
\begin{array}{c}
\Phi_n^+ \\
\Phi_n^- \\
\Phi_{n-1}^+ \\
\Phi_{n-1}^-
\end{array}\right),\,\,\,
\end{equation}
with
\begin{widetext}
\begin{equation*}
\hat{M}=
\left[ \begin{array}{cccc}
-t_{n,n+1}/\Delta_{n,n+1}& t_{n,n+1}^s/\Delta_{n,n+1}&0&0\\
t_{n,n+1}^s/\Delta_{n,n+1}&-t_{n,n+1}/\Delta_{n,n+1}&0&0\\
0&0&1&0\\
0&0&0&1
\end{array}\right]
\left[ \begin{array}{cccc}
(E-\varepsilon_n^+-\Sigma _n^+)& t_n^{so}&t_{n-1,n}&t_{n-1,n}^s\\
t_n^{so}&(E-\varepsilon_n^--\Sigma _n^-)&t_{n-1,n}^s&t_{n-1,n}\\
1&0&0&0\\
0&1&0&0
\end{array}\right]
\end{equation*}
\end{widetext}
and $\Delta_{n,n+1}=(t_{n,n+1})^2-(t_{n,n+1}^s)^2$. Assuming plane wave
functions for the electrons $\Phi_n=\sum_\sigma (A^\sigma
e^{ik_Lna}+B^\sigma e^{-ik_Lna})$ for $n\leq 0$ and $\Phi_n=\sum_\sigma
C^\sigma e^{ik_Lna}$ for $n\geq N+1$ in the left and right electrodes
respectively, we can express the output wave amplitude $C^\sigma$ in
terms of the input wave amplitude $A^\sigma$ and the transmission,
$$T^\pm(E)=\frac{|C^\pm|^2\sin(k_R^\pm a)}{|A^+|^2\sin(k_L^+ a)+
|A^-|^2\sin(k_L^- a)}.$$ We choose a normalized incident amplitude
$A^\pm=1/\sqrt{|\sin(k_L^\pm a)|}$. The net current primarily comes
from transmission of electrons of energy between the electrodes' Fermi
energies and is calculated as \cite{datt}
$$I=\frac{e^2}{h}\sum_\sigma \int_{-\infty}^\infty
dE\, T^\sigma (E)[f_L(E)-f_R(E)]$$ with the Fermi function
$f(E)=1/\exp[(E-E_F)/k_B T]$ and the room temperature $T=300$ K. For
ferromagnetic electrodes the megnetoresistance is defined as the
percentage change of resistance between parallel and anti-parallel
configurations $R_m=(R_{\text anti}-R_{\text paral})/R_{\text
anti}=(I_{\text paral}-I_{\text anti})/I_{\text paral}$.

\begin{figure}
\begin{center}
\begin{picture}(450,200)
\put(0,0){\includegraphics{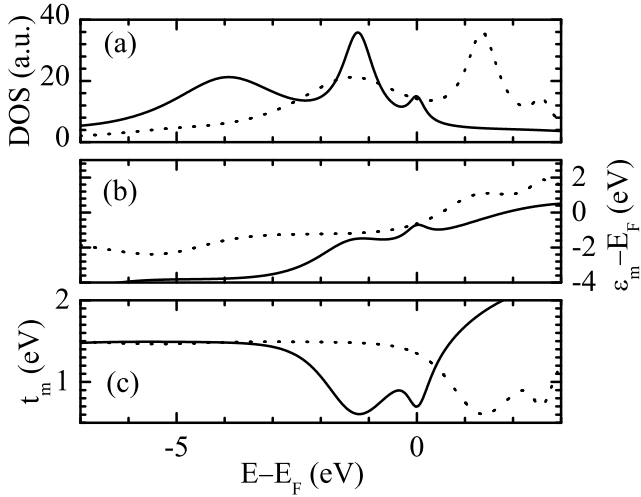}}
\end{picture}
\vspace{-0.8cm} 
\protect\caption{Energy dependence of the parameters
$\varepsilon_m$ (b) and $t_m$ (c) for ferromagnetic Fe. Solid curves
and dotted curves correspond to spin-up and spin-down electrons. The
resulting bulk DOS is also shown in (a).} \label{fig:fig2}
\end{center}
\end{figure}

In metals, an electron of energy $E$ may come from different energy
bands. The corresponding effective parameters $\varepsilon_{mX}^\sigma$
and $t_{mX}^\sigma$ are then an average of the parameters of these bands 
and are energy dependent. In linear or quasi-equilibrium system, they are
approximately the values near the Fermi energy. In a non-equilibrium
system, if the difference of the Fermi levels between the two
electrodes is comparable to the energy band width
of the metals, the energy dependence should be taken into account. In
the existing experiments the applied bias can be higher than 4 eV,
which is larger than the width of the $d$ bands where the Fermi level
locates in many metals. In the case of Ferromagnetic Fe which
exemplifies the electrode material here, approximately five bands can
be identified from the density of states (DOS) near the Fermi energy of
the bulk material \cite{sing}. For the spin-up (majority) electrons,
the five bands locate approximately at
2.5, 0, $-0.68$, $-3.4$, and $-7$ eV above the Fermi energy with band
width 6, 0.3, 0.6, 4.1, and 3.7 eV respectively.
For the spin-down (minority) electrons, the energy bands are the same
as above but shifted $2.58$ eV to higher energy. Using Lorentzian
broadening, we can mimic the bulk DOS and extract the parameters
$\varepsilon_m^\sigma$ and $t_m^\sigma$ as shown in
Fig.~\ref{fig:fig2}. At the Fermi energy, we get the hopping parameters
$0.39$ eV and $1.62$ eV for spin-up and spin-down electrons
respectively which coincide with the result obtained from the Fermi
velocity \cite{zwol}.

\begin{figure}
\begin{center}
\begin{picture}(450,220)
\put(0,0){\includegraphics{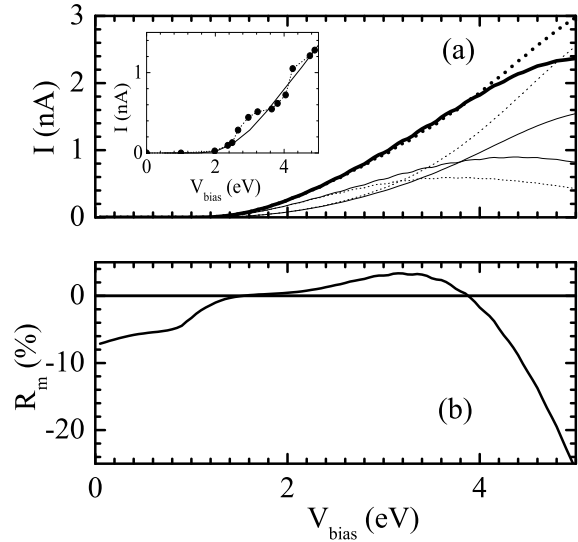}}
\end{picture}
\vspace{-0.8cm} 
\protect\caption{(a) The I-V curve of a 30-basepair Poly(G)-Poly(C) 
DNA between two ferromagnetic Fe electrodes of parallel (solid curves)
and anti-parallel (dotted) configurations. The thicker curves
illustrate the total currents and the thinner ones the contribution
from the two spin branches. (b) The magnetoresistance vesus the applied
bias potential. In the inset of (a), we show our theoretical fit (thick
solid curve) to the experimental result (filled circles connected by
dotted lines) in Ref.~\cite{pora1}. } \label{fig:fig3}
\end{center}
\end{figure}

Just as in Ref.~\cite{li}, we extract the parameters of the DNA chain
by fitting the experimental result of Ref.~\cite{pora1}. By evaluating
the energy-dependent parameters $\varepsilon_m^\sigma$ and $t_m^\sigma$
from Platinum's band structure \cite{roge}, we can fit the experimental
result as shown in the inset of Fig.~\ref{fig:fig3}(a). As a result, we
find that the hopping parameter is $t_d=0.6$ eV, the equilibrium Fermi
energy is 1.73 eV higher than the DNA HOMO on-site energy, the contact
parameters are $t_{dm}=0.019$ eV for the right electrode and $0.013$ eV
for the left, 1/3 of the bias voltage drop at the right contact and 2/3
at the left. The above parameters are close to those obtained in
Ref.~\cite{li} except for a larger $t_{dm}$ in the present case.

\begin{figure}
\begin{center}
\begin{picture}(420,170)
\put(0,0){\includegraphics{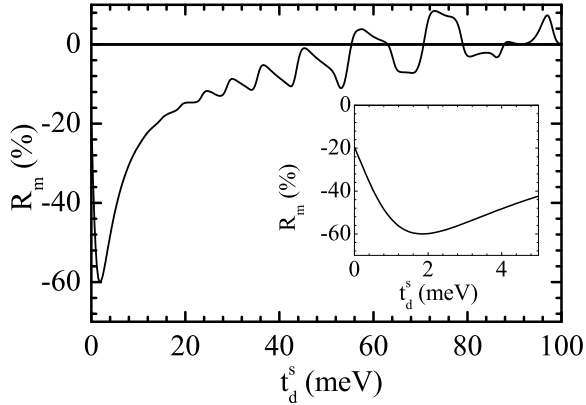}}
\end{picture}
\vspace{-0.8cm} 
\protect\caption{The magnetoresistance vesus the
spin-flip parameter $t_d^s$ in DNA when a bias of 4.8 eV is applied.
The other parameters are the same as in Fig.~\ref{fig:fig3}. }
\label{fig:fig4}
\end{center}
\end{figure}

For ferromagnetic Fe electrodes as shown in Fig.~\ref{fig:fig3}(a),
we get two I-V curves corresponding to the parallel (solid curves) and
antiparallel (dotted curves) magnetization configurations for the left
and right electrodes, using the same contact parameter $t_{dm}=0.02$ eV
for the two electrodes. In Fig.~\ref{fig:fig3}(b), the magnetoresistance
which describe the percentage change of the resistance of the system
when being switched from parallel to antiparallel configuration,
is plotted. In contrast to the results where constant values of the parameters
$\varepsilon_m^\sigma$ and $t_m^\sigma$ at the Fermi energy were used, here
we find a much smaller magnetoresistance until a strong bias is applied.
This can be understood from the energy dependence of $t_m$ shown in
Fig.~\ref{fig:fig2}(c). Instead of two parallel lines at 1.4 eV and 0.6 eV,
the curves of $t_m$ vs $E$ for spin-up and spin-down electrons cross
near the Fermi energy. This crossing makes the magnetoresistance disappear
around the bias voltage $V_{\rm{bias}}=2$ eV. The increase of magnetoresistance
at higher bias voltage results from the increasing $t_m$ spread between
spin-up and spin-down electrons in the range of 2 eV around the Fermi energy.

In Fig. \ref{fig:fig4}, we show how a spin-flip mechanism can affect
the spin injection, assuming that the spin-flip can happen only when an
electron jump from one site to another. The magnetoresistance does not
decay monotonically to zero. Instead, the magnetoresistance is enhanced
when there is a very weak spin-flip coupling ($t_d^s<1.9$ meV) as a
result of the quantum interference in the system. In the transmission
spectrum, peaks are slightly split with the increase of $t_d^s$,
indicating the mixing of the spin-up and spin-down states in the system
due to the spin-flip coupling. We observe an increase in
magnetoresistance from 20\% at $t_d^s=0$ to 60\% at $t_d^s=1.9$ meV as
displayed in the inset of Fig.~\ref{fig:fig4}. Then the
magnetoresistance decreases smoothly until $t_d^s=20$ meV. Above that
value, the magnetoresitance begins to oscillate when it decays to zero.

In summary, we have investigated the quantum spin transport through a
short DNA chain connected to ferromagnetic electrodes. We have used a
tight-binding model to describe the system where the parameters are
extracted from the experimental results and realistic energy bands of
metals. We find that the energy band structure of the ferromagnetic
electrodes significantly affects the resulting spin transport. In the
presence of the spin-flip mechanism, enhancement and oscillation of
magnetoresistance due to mixing of spin states are also observed.

The work of T.C. has been supported by the Canada Research Chair
Program and the Canadian Foundation for Innovation (CFI) Grant.

\end{document}